\documentstyle[aps,multicol,epsf]{revtex}

\begin{document}
\draft

\title{
Mesoscopics and fluctuations in networks
}

\author{
S.N. Dorogovtsev$^{1, 2, \ast}$ and A.N. Samukhin$^{1, 2, \dagger}$ 
}

\address{
$^{1}$ A.F. Ioffe Physico-Technical Institute, 194021 St. Petersburg, Russia\\
$^{2}$ Departamento de F\'\i sica and Centro de F\'\i sica do Porto, Faculdade 
de Ci\^encias, 
Universidade do Porto\\
Rua do Campo Alegre 687, 4169-007 Porto, Portugal
}

\maketitle

\begin{abstract}
We describe fluctuations in finite-size networks with a complex distribution of connections, $P(k)$. 
We show that the spectrum of fluctuations of the number of vertices with a given degree is Poissonian. These mesoscopic fluctuations are strong in the large-degree region, where $P(k) \lesssim 1/N$ ($N$ is the total number of vertices in a network), and are important in networks with fat-tailed degree 
distributions.  
\end{abstract}


\begin{multicols}{2}

\narrowtext


Fluctuations in finite systems, which vanish in the infinite system limit, is 
a basic topic of mesoscopic physics. 
The study of these fluctuations is an exciting field of condensed matter physics \cite{bookblw91,booky97}, where small objects attract much attention. 

Despite the fast recent progress in the statistical physics of networks 
\cite{w99,s01,ab02,dm02,bookdm02}, the finite-size effects in networks are surprisingly poorly studied. However, real networks, as a rule, are small objects, and their finiteness is a factor of primary importance. Impressive results obtained for infinite nets often turn out to be incorrect if we take into account the finiteness of networks (see detailed discussion in Refs. \cite{dm02,bookdm02}). For example, the absolute random-damage stability   
of infinite networks with fat-tailed degree distributions 
and the absence of an epidemic threshold in these nets (that is, their absolute disease vulnerability) 
vanish in finite-size networks \cite{ml01}. 

The finiteness of networks with fat-tailed degree distributions cutoffs the tails of the degree distributions (see, e.g., Refs. \cite{dms01,kr02,bck01}). 
In this communication we consider another mesoscopic effect which is clearly observable in empirical data. We describe fluctuations of the number of vertices $N(k)$ with a given number of connections $k$ in a network. $k$ is called degree, or, not quite rigorously, connectivity of a vertex. 
We consider a simple equilibrium situation, where the total numbers of vertices $N$ and edges $L$ in a network are constant, and comment a more complex case of fluctuations in growing networks (see calculations in Ref. \cite{kr02}). 

Our general results are formulated as follows: 
for any $k$, the fluctuations of $N(k)$ are described by the Poisson distribution. These fluctuations are of primary importance in the large-degree range, where the average $\langle N(k) \rangle \lesssim 1$. 
 
A random network is, actually, a statistical ensemble $G$ which includes numerous members $g \in G$ \cite{bck01,dms02}. These are different particular realizations of the network. Each member $g$ enters into the ensemble with its statistical weight $\Pi(g)$. 

When empirical researchers study a single realization of an ensemble, they measure $N(k)$, i.e. the number of vertices of a degree $k$ in this realization. 
If they have a possibility to investigate the complete ensemble, they 
can obtain the distribution of the fluctuations of $N(k)$ at any $k$, that is 
${\cal P}(N(k))$. In particular, the first moment of this distribution provides the degree distribution: $\langle N(k) \rangle = N P(k)$. 
So, studying only one member of the ensemble or a few ones, one can find the degree distribution only approximately. 

There is a specific network construction, where fluctuations of $N(k)$ are absent by definition.  
These are so called ``labeled random graphs with a given degree sequence''  
\cite{bbk72} from graph theory, which are, loosely speaking, the maximally random graphs with a given sequence $\{N(k)\}$, $\sum_k N(k)=N$. Consequently, in this situation, 
$P(k)=N(k)/N$. These graphs are extensively used in numerous applications 
\cite{mr95,cebh00,nsw00,n02}, and, of course, the absence of fluctuations is related only to their $N(k)$. So that, this is an exception. 

Here we study fluctuations in the dynamically constructed 
networks that are equivalent to 
the random graphs with a given degree sequence in the thermodynamic 
limit, where fluctuations vanish. Let us consider an equilibrium random network with a fixed number of vertices $N$ in which all vertices are statistically independent. 
Let a full set of probabilities $\{P(k)\},\ k=0,1,2,\ldots$ 
that a vertex has degree $k$ be given,  
but the total number of edges in particular members of this ensemble be not fixed. 
Taking into account the statistical independence of vertices we arrive at a standard combinatorial problem. 
So, 
a probability that a sequence $\{N(k)\}$ ($\sum_k N(k)=N$) is realized in a member of the ensemble has the following form:   
\begin{equation}
{\cal P}(\{N(k)\}) = N!  \prod_k \frac{[P(k)]^{N(k)}}{N(k)!}
\, .   
\label{e0}
\end{equation} 
This is a polynomial distribution from which the binomial form of the distribution of the numbers $N(k)$ (the fluctuation spectrum of $N(k)$) directly follows:  
\begin{equation}
{\cal P}(N(k)) = {N \choose N(k)} P(k)^{N(k)} [1 - P(k)]^{N-N(k)}
\, ,    
\label{e00}
\end{equation}     
This form, for sufficiently large networks, results in the Poisson spectrum of the fluctuations: 
\begin{equation}
{\cal P}(N(k)) = e^{-\langle N(k) \rangle}\, \frac{\langle N(k) \rangle^{N(k)}}{N(k)!} 
\, ,   
\label{e000}
\end{equation} 
which in turn approaches the Gaussian form 
\begin{equation}
{\cal P}(N(k)) \cong \frac{1}{\sqrt{2\pi\langle N(k) \rangle}}  
\exp\! 
\left\{\!-\frac{[(N(k) - \langle N(k) \rangle]^2}{2\langle N(k) \rangle}\right\}
\! .       
\label{e04}
\end{equation} 
when $\langle N(k) \rangle \gg 1$ (or $P(k) \gg 1/N$) while the ratio $[N(k) - \langle N(k) \rangle]/\sqrt{\langle N(k) \rangle}$ is fixed, that is when relative fluctuations are small.  
Note that these general results are valid both for directed and undirected networks. 

Now, let the number $L$ of edges in the realizations of an ensemble be fixed. 
In this event, the combinatorics at finite $N$ turns out to be slightly more complex, and we use a more convenient approach for sufficiently 
large networks.     
We consider fluctuations in the dynamically constructed equilibrium networks (with the fixed number of vertices and edges) that are equivalent to 
the above networks with a fixed $L$ (and to 
the random graphs with a given degree sequence) in the thermodynamic 
limit, where fluctuations vanish. Let us remind the construction 
and a standard statistical mechanics formalism \cite{dms02} which 
allows easy calculation of averages. 

\begin{list}{}{\leftmargin=17pt} 

\item[(1)]
The ensemble includes all graphs with a given numbers of vertices, $N$, and edges, $L$. 

\item[(2)]
The stationary statistical weights of the realizations are a limiting result of the following process. Random ends of randomly chosen edges are rewired to preferentially chosen vertices at a rate $f(k_i)$, where the rate depends on the degree $k_i$ of a target vertex. 

\end{list} 

The function $f(k)$ and the mean degree $\overline{k} = 2L/N$ determine the 
structure of this random network. (For brevity, here we consider undirected networks and do not discuss multiplication of edges. This will not influence our results.) The resulting statistical weights in this ensemble are products $\Pi(g) \propto \prod_{i=1}^N p(k_i)$, where   

\begin{equation}
p(k>0) = \prod_{q=0}^{k-1} f(q)
\, , \ \ \ 
p(0) = 1
\, .  
\label{e1}
\end{equation}   
At large $N$, the asymptotic expression for the partition function 
$Z=\sum_{g \in G} \Pi(g)$ of the ensemble is 
\begin{equation}
Z(\{p(\tilde k)\},N,L) \sim \left( \frac{\overline{k}}{e x_s^2} \right)^L [\Phi(x_s)]^N 
\, .  
\label{e2}
\end{equation}   
where  
\begin{equation}
\Phi(x) = \sum_k \frac{p(k)}{k!} x^k 
\, ,   
\label{e3}
\end{equation}   
and $x_s$ is given by the equation 
$\overline{k} = x_s \Phi'(x_s)/\Phi(x_s)$. 

With this partition function, the following standard statistical mechanics relations are valid:   
\begin{eqnarray}
& & C_1(k) = \langle N(k) \rangle = 
\frac{\delta \ln Z(\{p(\tilde k)\},N,L)}{\delta \ln p(k)} 
\, ,  
\nonumber
\\[5pt] 
& & C_2(k,k') = \langle N(k)N(k') \rangle - \langle N(k) \rangle\langle N(k') \rangle =  
\nonumber
\\[5pt] 
& & 
\phantom{WWWWWWWu}\frac{\delta^2 \ln Z(\{p(\tilde k)\},N,L)}{\delta \ln p(k) \delta \ln p(k')} 
\, , 
\label{e5}
\end{eqnarray}   
and so on for the higher cumulants $C_n(k_1, \ldots k_n)$ which contain higher order products of $N(k)$. 

Substituting Eqs. (\ref{e2}) and (\ref{e3}) into Eq. (\ref{e5}) readily shows that at sufficiently large $N$, when $\langle N(k) \rangle \ll N$,  

\begin{list}{}{\leftmargin=17pt}

\item[(1)]
$C_n(k_1, \ldots k_n)$ is nonzero only if $k_1 = k_2 = \ldots = k_n$, so there is no correlation between numbers of vertices of different degrees,  
and that 

\item[(2)]
all the cumulants are equal:  
$C_{n\geq 1}(k) = C_1(k) = \langle N(k) \rangle = N P(k)$. 

\end{list}
The latter is the characteristic property of the Poisson distribution. 
Consequently, the spectrum of the fluctuations of $N(k)$ is Poissonian, and we again arrive at the same expressions (\ref{e000}) and (\ref{e04}) as for networks with fluctuating $L$. This universal result does not depend on the specific parameters of the model which can be easily generalized. 


\begin{figure}
\epsfxsize=80mm
\centerline{\epsffile{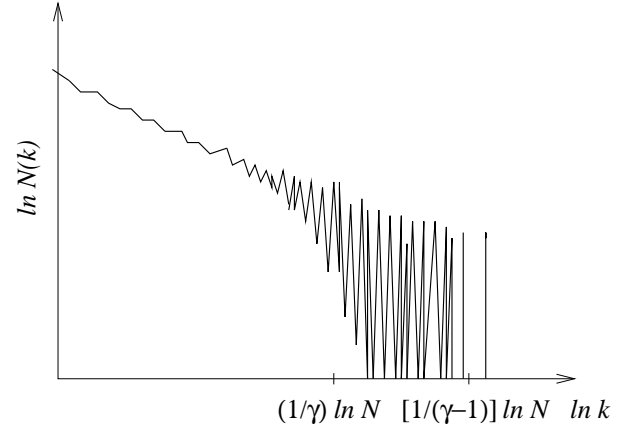}} 
\caption{
Typical (schematic) log-log plot of the number $N(k)$ of vertices of degree $k$ in a scale-free network (one realization). $\gamma$ is the exponent of the degree distribution. The fluctuations are observable in the region 
$(1/\gamma)\ln N  \protect\lesssim \ln k \protect\lesssim [1/(\gamma-1)]\ln N$.
}
\label{f1}
\end{figure}

  
The fluctuations of $N(k)$ vanish in the thermodynamic limit 
if $k$ is fixed. 
On the other hand, the fluctuations are strong in the region of large degrees where $\langle N(k) \rangle \lesssim 1$, that is $P(k) \lesssim 1/N$. 
The condition $P(k_m) = 1/N$ determines the lower boundary $k_m$ of this region. 
The position of the cut-off of the fat-tailed degree distribution may depend on many factors. In general terms, the maximum possible degree of the cut-off $k_c$ is determined by the finite size of the network: $N\int_{k_c}^\infty dk P_\infty (k) \sim 1$, where $P_\infty (k)$ is the degree distribution of the corresponding infinite network (note the discussion of the complex problem of the cut-off position in Ref. \cite{bck01}). $k_c$ is an estimate of the maximum vertex degree in a network and so is the upper boundary of the region, where the strong fluctuations are observable.   
In so-called scale-free networks, where $P_\infty (k) \propto k^{-\gamma}$, 
the strong fluctuations should be seen in the broad region $(1/\gamma)\ln N \lesssim \ln k \lesssim [1/(\gamma-1)]\ln N$ (see Fig. \ref{f1}). 

One should note that in equilibrium networks, the fat-tailed regime can be realized only starting from some threshold value of the mean degree \cite{dms02}. 
Below this threshold, a size-independent cut-off of the degree distribution sharply narrows the region of strong fluctuations. 

The Poisson spectrum of fluctuations of $N(k)$ in equilibrium networks is a natural consequence of the statistical independence of their vertices. 
In models of growing networks (e.g., see Refs. \cite{ba99,krl00,dms00}), the fluctuations of $N(k)$ were calculated only in a small degree region and were found to be Gaussian \cite{kr02}. The interesting region of strong fluctuations in growing nets (and in the Simon model \cite{s55}) is not studied yet. Although the problem is open, we suggest that the complete fluctuation spectrum of growing networks does not differ essentially from the spectrum of the fluctuations in equilibrium networks.  

In summary, we have shown that the fluctuation spectrum of the number of vertices of a given degree is Poissonian in equilibrium networks. These fluctuations are a prominent mesoscopic effect and are seen in the large-degree region of empirical degree distributions. As a rule, empirical researchers try to avoid fluctuations by passing to cumulative distributions. But fluctuations are interesting, and the simple fluctuations of $N(k)$ that we have considered is only a very particular type of fluctuations in networks. We hope to attract the attention of empirical researches to this problem. 
\\[-5pt]

S.N.D. was partially supported by the project POCTI/1999/FIS/33141.  
A.N.S. thanks the NATO program OUTREACH for support. 
Special thanks to the Centro de F\'\i sica do Porto and J.F.F.~Mendes.  
\\

\noindent
{\small $^{\ast}$      Electronic address: sdorogov@fc.up.pt} \\
{\small $^{\dagger}$ 
Electronic address: samukhin@fc.up.pt}

\end{multicols}

\end{document}